\documentclass[sigconf]{acmart}
\usepackage{multirow}
\usepackage{amsmath}
\usepackage{subfigure}
\usepackage{amsthm,amsmath,amssymb}
\usepackage{mathrsfs}
\usepackage{enumerate}
\usepackage{graphicx}
\usepackage{booktabs}
\usepackage{multirow}

\newcommand\Tstrut{\rule{0pt}{2.6ex}}

\usepackage[noend]{algpseudocode}

\usepackage{algorithmicx,algorithm}

\AtBeginDocument{%
  }

\copyrightyear{2022}
\acmYear{2022}
\setcopyright{acmcopyright}
\acmConference[SIGIR '22]{Proceedings of the 45th International ACM SIGIR Conference on Research and Development in Information Retrieval}{July 11--15, 2022}{Madrid, Spain.}
\acmBooktitle{Proceedings of the 45th International ACM SIGIR Conference on Research and Development in Information Retrieval (SIGIR '22), July 11--15, 2022, Madrid, Spain}
\acmPrice{15.00}
\acmISBN{978-1-4503-8732-3/22/07}
\acmDOI{10.1145/3477495.3532042}

\begin{document}
\fancyhead{}
\title{Preference Enhanced Social Influence Modeling for Network-Aware Cascade Prediction}

\author{Likang Wu}
\email{wulk@mail.ustc.edu.cn}
\orcid{0000-0002-4929-8587}
\affiliation{%
  \institution{Anhui Province Key Laboratory of Big Data Analysis and Application, School of Computer Science and Technology, University of Science and Technology of China \& State Key Laboratory of Cognitive Intelligence}
  \city{Hefei}
  \state{Anhui}
  \country{China}
}

\author{Hao Wang}\email{wanghao3@ustc.edu.cn}
\authornote{Corresponding Author.}
\orcid{0000-0002-4929-8587}
\affiliation{%
  \institution{Anhui Province Key Laboratory of Big Data Analysis and Application, School of Computer Science and Technology, University of Science and Technology of China \& State Key Laboratory of Cognitive Intelligence}
  \city{Hefei}
  \state{Anhui}
  \country{China}
}

\author{Enhong Chen}\email{cheneh@ustc.edu.cn}
\orcid{0000-0002-4835-4102}
\affiliation{%
  \institution{Anhui Province Key Laboratory of Big Data Analysis and Application, School of Computer Science and Technology, University of Science and Technology of China \& State Key Laboratory of Cognitive Intelligence}
  \city{Hefei}
  \state{Anhui}
  \country{China}
}

\author{Zhi Li}\email{zhili03@mail.ustc.edu.cn}
\affiliation{%
  \institution{Anhui Province Key Laboratory of Big Data Analysis and Application, School of Data Science, University of Science and Technology of China \& State Key Laboratory of Cognitive Intelligence}
  \city{Hefei}
  \state{Anhui}
  \country{China}
}

\author{Hongke Zhao}\email{hongke@tju.edu.cn}
\affiliation{%
  \institution{College of Management and Economics, Tianjin University}
  \city{Tianjin}
  \country{China}
}

\author{Jianhui Ma}\email{jianhui@ustc.edu.cn}
\affiliation{%
  \institution{Anhui Province Key Laboratory of Big Data Analysis and Application, School of Computer Science and Technology, University of Science and Technology of China \& State Key Laboratory of Cognitive Intelligence}
  \city{Hefei}
  \state{Anhui}
  \country{China}
}
\renewcommand{\shortauthors}{Trovato et al.}

\begin{abstract}
Network-aware cascade size prediction aims to predict the final reposted number of user-generated information via modeling the propagation process in social networks. Estimating the user's reposting probability by social influence, namely state activation plays an important role in the information diffusion process.
Therefore, Graph Neural Networks (GNN), which can simulate the information interaction between nodes, has been proved as an effective scheme to handle this prediction task. 
However, existing studies including GNN-based models usually neglect a vital factor of user's preference which influences the state activation deeply.
To that end, we propose a novel framework to promote cascade size prediction by enhancing the user preference modeling according to three stages, i.e., preference topics generation, preference shift modeling, and social influence activation. Our end-to-end method makes the user activating process of information diffusion more adaptive and accurate. Extensive experiments on two large-scale real-world datasets have clearly demonstrated the effectiveness of our proposed model compared to state-of-the-art baselines.
\end{abstract}




\maketitle

\section{Introduction}
On popular social media platforms, e.g., Facebook, Twitter, Sina Weibo, etc, tens of millions of User-Generated online Content (UGC) are spread through social networks every day. With such a vast amount of information, predicting the cascade size of propagating information, i.e., the final reposted number, is valuable for people out of the dilemma of information explosion, such as discovering the hot topic in advance~\cite{NaseriZ19}.

To predict the future cascade size as accurately as possible, the researchers need to consider the structural information of the social network where messages are propagated. The early diffusion-based methods~\cite{wang2015learning,ohsaka2017coarsening} make a strong assumption that the underlying diffusion follows a known prior distribution, which is usually not very appropriate for actual cascade prediction. To incorporate the network topology, feature-based methods design many effective features including cascade density~\cite{guille2012predictive}, node degree~\cite{lerman2008analysis}, the number of leaf nodes of local graph~\cite{guo2016comparison}, community features~\cite{clauset2004finding} and so on. Unfortunately, the performance of such methods heavily depends on the quality of hand-crafted features that are generally extracted heuristically. And the user's state activation affected by social influence cannot be simulated by them, which is crucial to judge a user's reposting probability for a message. Recently, with the booming of graph neural networks for network embedding~\cite{kipf2016semi,velivckovic2017graph,hamilton2017inductive,wu2021learning,qiao2019structure}, such network-aware predictions via representation-based models can effectively model the interpersonal influence~\cite{cao2020popularity,liu2021content} between nodes (users) and capture the network structure~\cite{chen2019information,li2017deepcas,wang2019information,wu2022estimating} at the same time, which overcome the technical obstacle of heuristic features and gradually make more and more breakthroughs in the area of information cascade prediction.

However, most of the existing network-aware methods including representation-based models ignore the importance of personal preference in the activation process of user's sharing state. When considering the social impact, although users will receive a lot of messages on social platforms, they usually only repost the content according to their personal preference. What's more, unlike the static node (paper) of information (citation) cascade in scholar networks, the users' preferences will show dynamic drift in different periods in a social media platform. For instance, a reader of Agatha's novel often shares some novels' comments and discussions, but she may be fascinated by suspense films after watching the release of the new version of~\textit{Murder on the Oriental Express} and begin to follow and retweet some tweets of relevant movies. Therefore, it is necessary for the cascade-prediction model to capture the user preference and its dynamism, which promotes the accuracy of activation prediction in complex diffusion processes.

Unfortunately, the exploration of user preference modeling in information cascade prediction is largely limited with great challenges. First, it is hard to estimate the preference of a user and generate a corresponding representation vector for the prediction task. Although there is a study~\cite{liu2021content} that utilizes the topic model to mine preferred topics for user groups in the social media platform, whose non-end-to-end learning framework inevitably leads to a deviation between the generated preference topics and the actual task. Second, the evolution of dynamic preference has not been explored in the area of network-aware information diffusion, we are supposed to design a stable and effective pattern-capturing approach that is able to embed into the overall model.

To solve the dilemmas mentioned above, in this paper, we present a focused study on the framework of cascade size prediction from a novel user preference intensive perspective. Specifically, underlying topics of information cascade and history retweets of users over the graph are first generated via a text content reconstructing process by the neural topic model~\cite{srivastava2017autoencoding}, where each of these topics is modeled as a probability distribution over a fixed vocabulary. With the generated topic representation, we utilize two different paradigms to learn the long-term preference and compare their performances, i.e., Bi-LSTM~\cite{li2018learning}, Attentive Asymmetric-SVD~\cite{yu2019adaptive}. The evolution trend of short-term preference is captured by the sequential model LSTM. 
Finally, we carefully design a GNN-based state-activating module to characterize the underlying social structure, where node embeddings incorporate the learned preference, and neighborhood aggregation strategy can effectively capture the inter-influence from social networks.
We verify the effectiveness of PEG (our proposed user Preference-Enhanced Graph model) on two real-world data in Sina Weibo and Tweet. Experimental results demonstrate that our proposed method significantly outperforms all the state-of-the-art methods.

\section{Problem and Methodologies}
\begin{figure}[t]
  \includegraphics[width=0.44\textwidth]{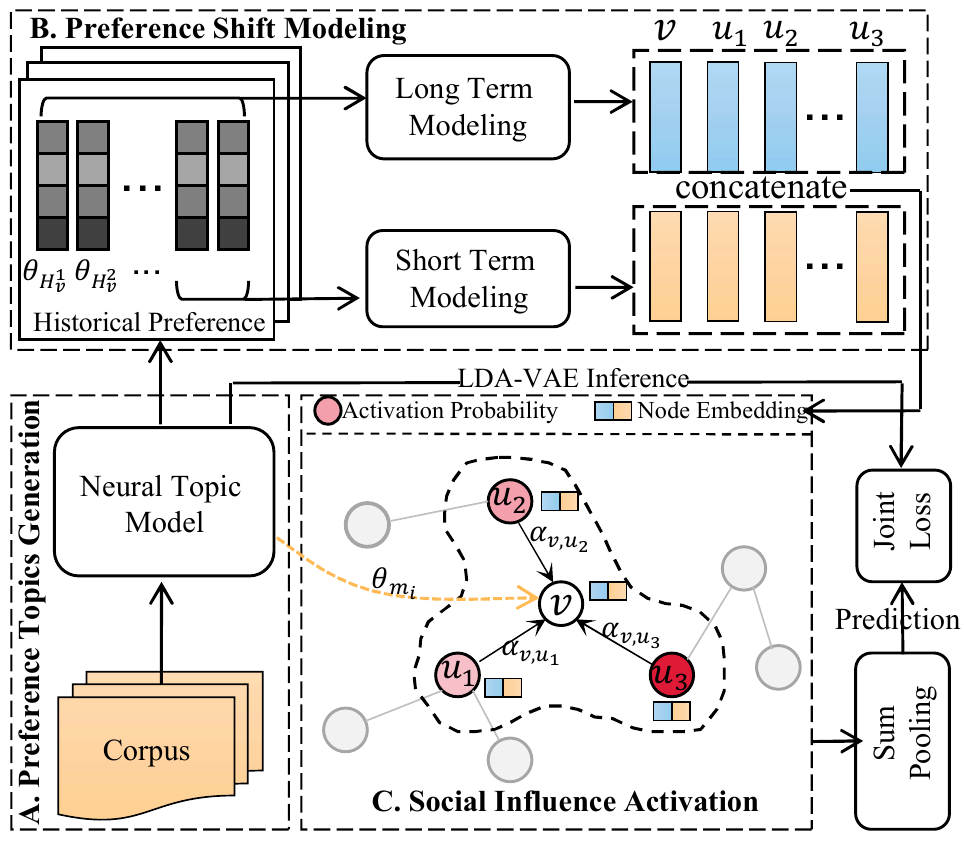}
  \caption{Framework of our PEG model. The darker the node's color, the larger the activation probability.}
  \label{fig:fk}
\end{figure}

In this section, we give the formal definition of the network-aware cascade size prediction and introduce the details of our proposed model. The framework of PEG is presented in Figure~\ref{fig:fk}.
\subsection{Problem Definition}
Our task can be viewed as a regression problem for user-generated information, and the goal is to predict the final number of reposted users in the social network. We formally define the task as:

\textbf{Network-aware cascade size prediction}. Given $N$ information items $\{m_1, m_2, ..., m_N\}$ and the corresponding active users list $C^{m_i}_T = \{u_1, u_2, ..., u_{n^{m_i}_T}\} $ who have been taken action on $m_i$$(i \in N)$ in the defined early observation window $T$. For the known social network $G = (V, E)$, $V$ is the set of all users, and $E \subseteq V \times V$ denotes the following relationships. The purpose is to predict the final reposted number $S_i$ of the information item $m_i$ in the future.

Compared with conventional cascade size prediction, the network-aware prediction problem~\cite{cao2020popularity,liu2021content} emphasizes the role of the following network structure within information diffusion, i.e., there are interactions between early adopters and potential active users, or among the potential active users. Capturing the cascading effect along the social network provides a more comprehensive context for the future popularity prediction of online content.

\subsection{Preference Topics Generation}
For a information cascade, semantic components of this UGC may trigger user's reposting behaviors, e.g., sports news attract significantly more attention than normal content for sports fans. Given the historical reposting list $\mathcal{H}_v = \{\mathcal{H}_v^1, \mathcal{H}_v^2, ..., \mathcal{H}_{v}^{n_v} \}$ of each user $v \in V$, we input the text of $N$ pieces of $m_i$ and $\{\mathcal{H}_v | v \in V\}$ as a corpus into the topic model to generate topic proportion vectors. To prevent the deviation between the generated topics and the actual task, we implement a neural topic module to be trained with the entire model. 
We appeals to LDA-VAE~\cite{srivastava2017autoencoding} to infer latent semantics underlying the corpus. Like general topic algorithms, for clarity, we present the generative process of text contents:
\begin{itemize}
\item  For each topic $k$:

$\triangleright$ Draw the topic state vector $\phi_{k} \sim \mathcal{N}\left(0, \epsilon^{2} I\right)$
\item For each text $t$ in the corpus:

$\triangleright$ Draw the topic proportion vector $\theta_{t} \sim \operatorname{Dir}\left(\operatorname{softmax}\left(\Phi\right)\right)$

$\triangleright$ For each word $w_{ti}$:

\quad $-$ Draw the topic assignment $z_{t i} \sim \operatorname{Multi}\left(\operatorname{softmax}\left(\theta_{t}\right)\right)$

\quad $-$ Draw the word $w_{t i} \sim \operatorname{Multi}\left(\beta_{z_{t i}}\right)$
\end{itemize}

The motivation of topic model is that all texts are generated by a mixture of $K$ latent topics. A fixed vocabulary $\mathcal{V}$ is captured from its corpus, and each text consists of a bag of words $\boldsymbol{w}_{t}=\left\{w_{1}, \ldots, w_{n}\right\}$ that comes from $\mathcal{V}$. The $k$-th topic is denoted by a probability distribution $\beta_{k} \in {\mathbb R}^{|\mathcal{V}|}$ of words over the vocabulary, $\beta=\{\beta_{k} | k=1,2,...,K\}$. For each text $t$, $\theta_{t} \in {\mathbb R}^{K}$ indicates its corresponding topics proportion, and we use the set of state vectors of topics $\Phi = \{ \phi_{k} \in \mathbb{R}^{d_{\phi}} | k=1,2,...,K \}$ to generate $\theta_{t}$. According to~\cite{srivastava2017autoencoding}, we aim to train a parametric variational model $q(\theta | \mathbf{w}) \approx p(\theta | \mathbf{w})$, all parameters are trained by optimizing the ELBO loss as follow:
\begin{small}
\begin{equation}
\mathcal{L}(q, p)=\mathbb{E}_{\theta \sim q}[\log p(w \mid \theta)]-D_{K L}(q(\theta \mid w) \| p(\theta)),
\label{eq:elbo1}
\end{equation}
\end{small}
where the subscript $t$ is omitted and $\theta$ represents the per-document topic distribution for the sake of simplicity.

\subsection{Preference Shift Modeling}
The preference of a user in the social network would present a dynamic shift, which is used to estimate her diffusion probability. In our work, we capture stable preferences in the long-term period and recent preference evolution in the short-term period, respectively.

We propose two learning paradigms to capture the long-term preference, i.e., Bi-LSTM~\cite{li2018learning}, Attentive Asymmetric-SVD~\cite{yu2019adaptive}.

\subsubsection{Bidirectional LSTM}
In our work, to obtain a global representation of historical preferences with fewer fluctuations, we draw support from the ability of Bi-LSTM that learns contextual long-term representation in both forward and backward directions. For user $v$, at each time step $i$ of the historical reposting list, the forward layer updates hidden state $\overrightarrow{h_v^{i}}$ with both the previous hidden state $\overrightarrow{h_v^{i-1}}$ and the current reposted item $\theta_{\mathcal{H}_v^{i}}$ ($\theta_{\mathcal{H}_v^i}$ indicates the topic proportion vector of the $i$-th historical retweet of user $v$,) ; while the backward layer with hidden state $\overleftarrow{h_v^{i}}$ is trained based on the future state $\overleftarrow{h_v^{i+1}}$ and the current item $\theta_{\mathcal{H}_v^{i}}$. Along this line, each hidden representation $h_v^{i}$ is presented as the concatenation of the forward state and backward state, i.e., $h_v^{i}= [ \overrightarrow{h_v^{i}} \| \overleftarrow{h_v^{i}} ]$ ($\|$ indicates the concatenation operation). After that, we can generate the unified representation of long-term preference $P_v^L$ for user $v$ through an average pooling layer:
$P_v^L=\operatorname{average}\left(h_v^{1}, h_v^{2}, \ldots, h_v^{|\mathcal{H}_v|}\right)$.

\subsubsection{Attentive Asymmetric-SVD}
We refer to the non-sequence-based Attentive ``Asymmetric-SVD'' paradigm, which represents the long-term effect through the items that the user interacts with:
\begin{small}
\begin{equation}
P_v^L={\sum}_{i=1}^{|\mathcal{H}_v|} \alpha_{v}^{i} h_v^{i} ,
\label{eq:pu}
\end{equation}
\end{small}
where $\alpha_{v}^{i}$ is the weighting score for item $i$ of user $v$, $h_v^{i} = \sigma (W_h \theta_{\mathcal{H}_v^i} + b_h)$, $W_h \in \mathbb{R}^{d_L \times K}$, $\sigma$ is the sigmoid activation function. It is reasonable to presume that not all items contribute equally, hence we assign higher (lower) weights for the corresponding items that are more informative (less informative) as:
\begin{small}
\begin{equation}
\begin{gathered}
\alpha_{v}^{i}=\frac{\exp \left(x_v^{i} \right)}{\sum_{j=1}^{|\mathcal{H}_v|} \exp \left(x_v^{j} \right)},~ ~ x_v^{i}=\sigma \left(W_{x} h_v^{i} +b_{x}\right) ,
\end{gathered}
\end{equation}
\end{small}
where $W_{x} \in \mathbb{R}^{d_L}$. Unlike Bi-LSTM, Attentive Asymmetric-SVD is a feedforward neural network with less computational cost.

For the short-term effect, we develop a LSTM rather than feedforward model to learn recent preference evolution, since the sequential model can explore the trend of dynamic preference to reveal the users’ present sharing motivations. Therefore for user $v$, the short-term preference is represented as $P_v^S = LSTM([\theta_{\mathcal{H}_{v}^{n_{v}-\tau+1}}, \theta_{\mathcal{H}_{v}^{n_{v}-\tau+2}}, \\ ..., \theta_{\mathcal{H}_{v}^{n_{v}}} ])$, where $\tau$ denotes the defined length of short-term session and $P_v^S \in \mathbb{R}^{d_S}$ is the hidden state of the last step.

\subsection{Social Influence Activation}
In social networks, the embedding of each user (node) $v$ is initialized as the fusion of preference vector $e_v = [P_{v}^{L}||P_{v}^{S}]$, the influence from social network on a user can be obtained by the neighborhood aggregation strategy of the graph neural network. With the message passing, the influence of active users further spreads to other users along with the following network. Considering the different types of relationships, the attention coefficient $\alpha_{v,u}$ in GAT~\cite{velivckovic2017graph} is used to represent the influence weight from user $u$ to user $v$.
Therefore, we define the influence of user $v$ from the social network as follow:
\begin{small}
\begin{equation}
\mathcal{I}_{v}^{(l)}=\left[ {\sum}_{u \in \mathcal{N}(v)} s_{u}^{(l)} \alpha_{v,u}^{(l)} W^{(l)} e_{u}^{(l)} ~\|~ \Gamma \left({\sum}_{u \in \mathcal{N}(v)} s_{u}^{(l)}\right) \theta_{m_i} \right],
\end{equation}
\end{small}
where $s_{u}^{(l)} \in \mathbb{R}$ is the updated activation probability of user $u$ at the $l$-th layer of message passing, whose updating method is shown in Eq. (\ref{eq:sv}). $W^{(l)} \in$ $\mathbb{R}^{d_g^{(l)} \times d_g^{(l)}}$ is a weight matrix, $e_{u}^{(l)} \in \mathbb{R}^{d_g^{(l)}}$ is the user $u$ 's embedding
at the $l$-th layer. $\theta_{m_i} \in \mathbb{R}^{K}$ is the topic proportion vector of $m_i$, $\|$ represents the concatenate operation, and the function $\Gamma(x) = 1$ if $x>0$, else 0. Then we obtain the user's embedding vector of the $(l+1)$-th layer as follow:
\begin{small}
\begin{equation}
e_{v}^{(l+1)}=\sigma\left(\xi^{(l)} W^{(l)} e_{v}^{(l)}+ (1-\xi^{(l)})W_{\mathcal{I}}^{(l)} \mathcal{I}_{v}^{(l)}\right),
\end{equation}
\end{small}
where $\sigma$ is the sigmoid function, $\xi^{(l)} \in \mathbb{R}$ is a weight parameter, $W_{\mathcal{I}}^{(l)} \in \mathbb{R}^{d_g^{(l)} \times\left(d_g^{(l)}+K\right)}$ is a weight matrix. 

During the message passing of GNN, we update the activation state of each node simultaneously via a gated mechanism:
\begin{small}
\begin{equation}
s_{v}^{(l+1)}=\sigma\left((1-\eta^{(l)}) s_{v}^{(l)}+\eta^{(l)} \left(\sum_{u \in \mathcal{N}(v)} {W_s^{(l)}}^{\top}\left[W^{(l)} e_{u}^{(l)} \| W^{(l)} e_{v}^{(l)}\right] s_{u}^{(l)}\right)\right),
\label{eq:sv}
\end{equation}
\end{small}
where $\sigma$ is the sigmoid activation function, $s_{v}^{(l+1)}$ denotes the updated activation probability of user $v$ at the $(l+1)$-th layer, $W_s^{(l)} \in \mathbb{R}^{2 d_g^{(l)}}$ is a weight matrix, and $\eta^{(l)} \in \mathbb{R}$ is the weight parameter to control the updating degree. The initial activation probability of the sharing user in the observation cascade is set to 1, otherwise 0.
\subsection{Prediction and Model Optimization}
The activation probability of each user is $s^{(L)} \in[0,1]$ after $L$ GNN layers. We calculate the size of the information cascade by the sum of all users' activation probability:
\begin{small}
\begin{equation}
\tilde{S}_{i}={\sum}_{u \in V} s_{u}^{(L)}.
\end{equation}
\end{small}

Here we use the Mean Relative Square Error (MRSE) loss $\mathcal{L}_m$ to optimize our model, which is designed as:
\begin{small}
\begin{equation}
\mathcal{L}_m=\frac{1}{N} {\sum}_{i=1}^{N}\left(\frac{\tilde{S}_{i}-S_{i}}{S_{i}}\right)^{2},
\end{equation}
\end{small}
where $N$ is the total number of information cascades, $S_{i}$ is the ground truth and $\tilde{S}_{i}$ is the final predicted cascade size. Besides, following~\cite{srivastava2017autoencoding}, the optimization loss for Eq. (\ref{eq:elbo1}) is computed as:  
\begin{small}
\begin{equation}
\begin{split}
\mathcal{L}_{\tau} = - \frac{1}{|\mathcal{D}|} \sum_{t \in \mathcal{D}}\left[\mathbb{E}_{\varepsilon \sim \mathcal{N}(0, I)}\left[\boldsymbol{w}_{t}^{\top} \log \left(\sigma\left(\mu_{t}+\varepsilon \Sigma_{t}^{1 / 2}\right)^{\top} \sigma(\beta)\right)\right]\right. \\
\left.-\frac{1}{2}\left(\operatorname{Tr}\left(\Sigma_{t}^{-1} \Sigma_{t}\right)+\left(\mu_{t}^{\prime}-\mu_{t}\right)^{\top} \Sigma_{t}^{\prime-1}\left(\mu_{t}^{\prime}-\mu_{t}\right)+\log \frac{\left|\Sigma_{t}^{\prime}\right|}{\left|\Sigma_{t}\right|}\right)\right],
\end{split}
\label{loss:topic}
\end{equation}
\end{small}
where the two terms in the $\sum (\cdot)$ correspond to the two terms in Eq. (\ref{eq:elbo1}), respectively. When computing the first term $\mathbb{E}_{\theta \sim q}[\log p(w | \theta)]$ of Eq. (\ref{eq:elbo1}), we approximate the variable $\theta \sim q(\theta | \boldsymbol{w})$ by reparameterizing function $\theta=\sigma\left(\mu+\varepsilon \Sigma^{1 / 2}\right)$ with the new variable $\varepsilon$, where $\sigma = \frac{1}{1 + e^{-x}}$, mean  $\mu=\operatorname{MLP}_{\mu}(\boldsymbol{w})$, variance  $\Sigma=\operatorname{MLP}_{\Sigma}(\boldsymbol{w})$, and $\varepsilon \sim \mathcal{N}(0, I)$. For the second term, we transform the Dirichlet distribution of $p(\theta)$ by the effective reparameterization function of Laplace approximation~\cite{srivastava2017autoencoding} with a logistic normal distribution $\mathrm{log} \mathcal{N}\left(\mu^{\prime}, \Sigma^{\prime}\right)$. Specifically, $\mu_{k}^{\prime}=\log \alpha_{k}-\frac{1}{K} \sum_{i}^K \log \alpha_{i}, \Sigma_{k k}^{\prime}=\left(1-\frac{2}{K}\right) \frac{1}{\alpha_{k}}+\frac{1}{K^{2}} \sum_{i}^K \frac{1}{\alpha_{i}}$, where  $\alpha=\sigma\left(\Phi\right)$ introduces the topic state vector. 
Finally, the joint loss $\mathcal{L} = \mathcal{L}_m + \lambda \mathcal{L}_{\tau}$ optimizes our entire end-to-end model and $\lambda$ is a trade-off parameter.

\section{Experiments}
In our experiments, we compare the \textbf{prediction performance} of our proposed PEG model with several sate-of-the-art methods on the real-world Sina Weibo and Twitter datasets. \textbf{Ablation studies} and \textbf{hyperparameter tune-up} experiments are also conducted.

\subsection{Datasets and Baselines}
\subsubsection{Sina Weibo} Sina Weibo~\footnote{https://www.aminer.cn/Influencelocality} is the most popular microblogging platform in China. Following~\cite{cao2020popularity}, we randomly select some seed users to obtain their information paths and following relationships from the global network. The final network contains 16,924 users, 1,226,478 edges and 3,583 information cascades. The average length of captured historical retweet is about 24. We set the observation time window to $T = 1.5 h$ and $T = 3 h$.
\subsubsection{Twitter}
There is no complete following network among users in the Twitter~\footnote{https://www.dropbox.com/s/7ewzdrbelpmrnxu/rumdetect2017.zip} dataset. Following the operation of~\cite{liu2021content}, we extract a total of 3,260 users and 12,752 edges and 2,025 information cascades. We set the observation time window to $T = 2 h$. Note that, we only implement and evaluate the preference topic generation module and the social influence activation module on Twitter, since this dataset does not capture retweeting history for users.

\subsubsection{Baselines}
There are not so many following network-aware prediction models for our task and we refer to the settings of previous works~\cite{cao2020popularity,liu2021content}.
We compare our proposed model with the following baselines, i.e., the Feature-Based model~\cite{cao2020popularity,liu2021content,wu2020multi}, the self-excited process generation model SEISMIC~\cite{zhao2015seismic}, the graph-based models including DeepCas~\cite{li2017deepcas}, CoupledGNN~\cite{cao2020popularity}, and TSGNN~\cite{liu2021content}. For the Feature-Based model, we use a logistic regression classifier that inputted several important heuristic features, e.g., the degree distribution of the network, the number of leaf nodes, etc.


\begin{table*}[t]
\large
\caption{Performance comparison on the datasets, where $^*$ indicates the best result among baselines.}
\centering
\resizebox{0.69\textwidth}{!}{
\begin{tabular}{c|cccc|cccc|cccc}
\hline
\textbf{Dataset} & \multicolumn{8}{c|}{Sina Weibo} & \multicolumn{4}{c}{Twitter}\\
\hline
\textbf{Observation} & \multicolumn{4}{c|}{1.5h} & \multicolumn{4}{c|}{3h} & \multicolumn{4}{c}{2h}\\
\hline
\textbf{Metrics} & MRSE & mRSE & MAPE & WP(\%) & MRSE & mRSE & MAPE & WP(\%) & MRSE & mRSE & MAPE & WP(\%)\\
\hline
\Tstrut Feature-Based & 0.2445 & 0.1866 & 0.3932 & 33.84 & 0.1650 & 0.0748 & 0.3062 & 18.83 & 0.1337 & 0.1002 & 0.3209 & 17.94 \\
SEISMIC & - & 0.2029 & -  & 37.54 & - & 0.0776 & -  & 22.45 & - & 0.0943 & -  & 17.51\\
DeepCas & 0.1933 & 0.1482 & 0.3655 & 29.01 & 0.1329 & 0.0538 & 0.2773 & 15.44 & 0.0969 & 0.0602 & 0.2705 & 14.28\\
CoupledGNN & 0.1840 & 0.1306$^*$ & 0.3598$^*$ & 28.31 & 0.1287 & 0.0473 & 0.2711 & 13.86 & 0.0874$^*$ & 0.0566 & 0.2519$^*$ & 12.70$^*$\\
TSGNN & 0.1825$^*$ & 0.1339 & 0.3610 & 28.06$^*$ & 0.1248$^*$ & 0.0455$^*$ & 0.2671$^*$ & 13.80$^*$& 0.0888 & 0.0547$^*$ & 0.2552 & 12.99\\
\hline
\Tstrut PEG-T & 0.1724 & 0.1197 & 0.3520 & 27.77 & 0.1165 & 0.0416 & 0.2593 & 13.42 & - & - & - & -\\
PEG-S & 0.1697 & 0.1205 & 0.3520 & 27.83 & 0.1139 & 0.0394 & 0.2588 & 13.31 & - & - & - & -\\
PEG-L & 0.1735 & 0.1226 & 0.3543 & 27.64 & 0.1150 & 0.0428 & 0.2601 & 13.57 & - & - & - & -\\
PEG-D & 0.1769 & 0.1240 & 0.3538 & 27.91 & 0.1184 & 0.0433 & 0.2629 & 13.72 & \bf 0.0859 & \bf 0.0523 & \bf 0.2480 & \bf 12.45\\
PEGA & 0.1692 & 0.1181 & 0.3529 & \bf 27.20 & 0.1136 & 0.0390 & \bf 0.2563 & \bf 12.97 & - & - & - & -\\
PEGL & \bf 0.1668 & \bf 0.1163 & \bf 0.3504 & 27.53 & \bf 0.1119 & \bf 0.0385 & 0.2579 & 13.35 & - & - & - & -\\
\hline
\end{tabular}}
\label{table:compar}
\end{table*}

\subsection{Implementation Details}
We optimize the parameters of our model with Adam~\cite{DBLP:journals/corr/KingmaB14}. Here we set $d_{\phi}=32$ for the topic state vector $\phi_{k}$, the vocabulary size $|\mathcal{V}| = 1074$ after removing low-frequency words (less than 30 occurrences) and meaningless words (prepositions, conjunctions, typos, etc). $d_L, d_S = 32$ is the dimension of hidden state of LSTM, $d_g^{(l)} = 64$ in each GNN layer, the trade-off parameter $\lambda = 0.1$ achieves the best performance, and the learning rate $\sim$ loguniform~$[e-8, 1]$ is optimized by Hyperopt~\cite{bergstra2013hyperopt}. Like GraphSage~\cite{hamilton2017inductive}, we train 64 information cascades and related local social network in each batch. The proportion of training/validation/testing is set as 70/10/20\%. The hyperparameter of topic numbers $K$, GNN layers $L$, and short-term length $\tau$ will be tested in our hyperparameter tune-up experiments.

\subsection{Results Analysis}
\subsubsection{Performance Comparison and Ablation Studies} We report the prediction results of our model and comparison methods in Table~\ref{table:compar}. Following~\cite{cao2020popularity,liu2021content}, we select several evaluation metrics, i.e., Mean Relative Square Error (MRSE), Median Relative Square Error (mRSE), Mean Absolute Percentage Error (MAPE),  Wrong Percentage Error (WP, error tolerance is 0.5). The PEG-based results in the lower half of Table~\ref{table:compar} present the ablation studies of our framework. PEG-T denotes PEG without topic generation (using tweet contents as the input of preference modeling module), PEG-D is PEG without preference shift modeling, PEG-S(-L) is that PEG only removes the short-term (long-term) modeling, PEGL and PEGA are that PEG uses Bi-LSTM and Attentive Asymmetric-SVD to model the long-term effect, respectively. Except for PEGA, other PEG's variants all use Bi-LSTM to capture the long-term preference. As shown in Table~\ref{table:compar}, PEG-based models achieve the best performance in four metrics on all experimental settings, which shows the efficiency of our learning paradigm. PEG-T outperform all baselines, which illustrates the necessity of capturing preference shift to model state activation. PEG-D obtains better scores than the most powerful topic-manner baseline TSGNN, which is in line with our intention of designing the end-to-end topic generation module. The better score of PEGL than both PEG-S and PEG-L shows that the fusion of long and short-term effects could promote the model's predictability. Besides, we can find that PEGL is slightly better than PEGA in our task. However, we recommend using PEGA in practical applications, because its good parallel ability can produce similar performance.
\subsubsection{Hyperparameter Tune-up} 
\begin{figure}[t]
  \centering
  \subfigure[K: number of topics]{
    \label{fig:K} 
    \includegraphics[width=0.31\columnwidth]{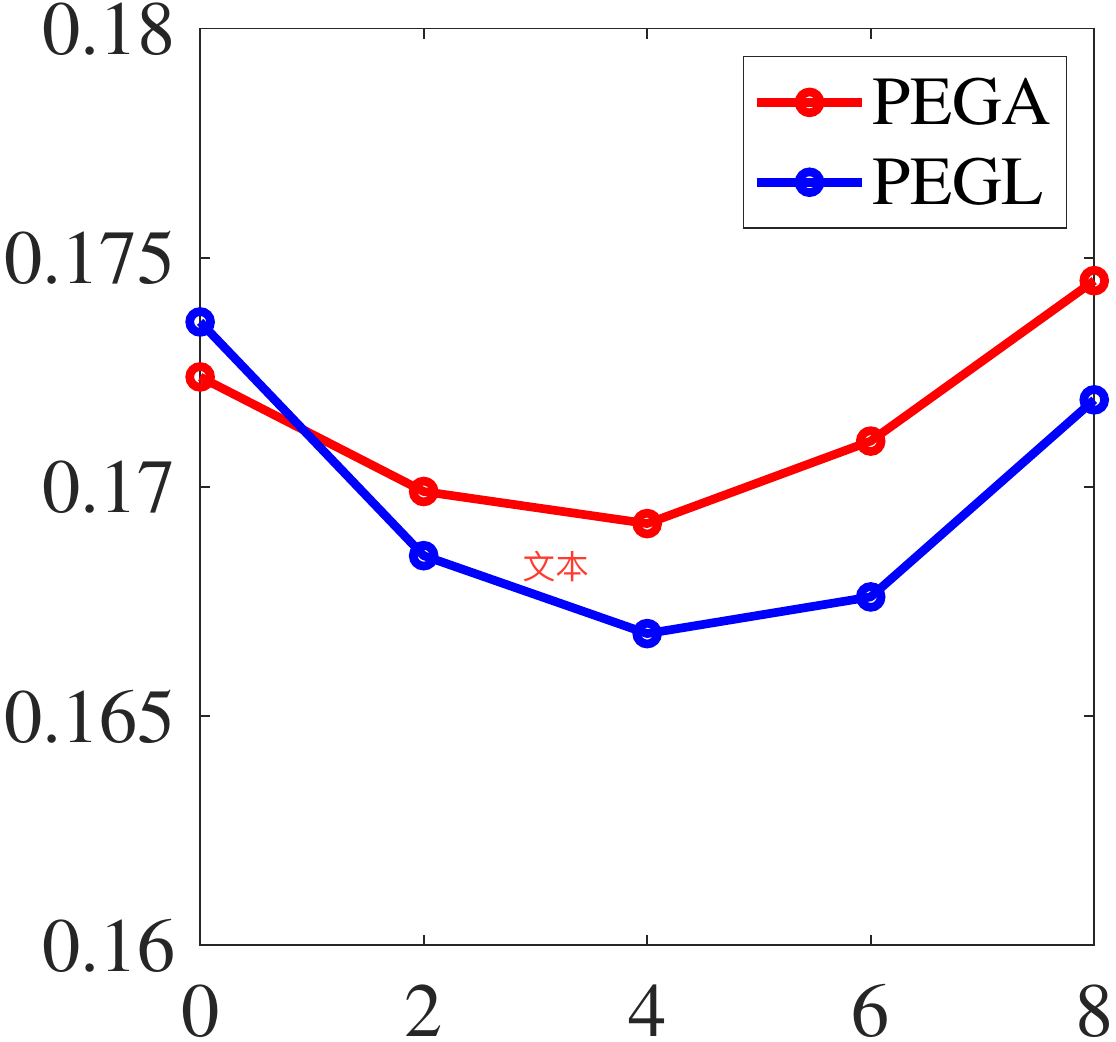}}
  \subfigure[L: GCN layers]{
    \label{fig:L} 
    \includegraphics[width=0.31\columnwidth]{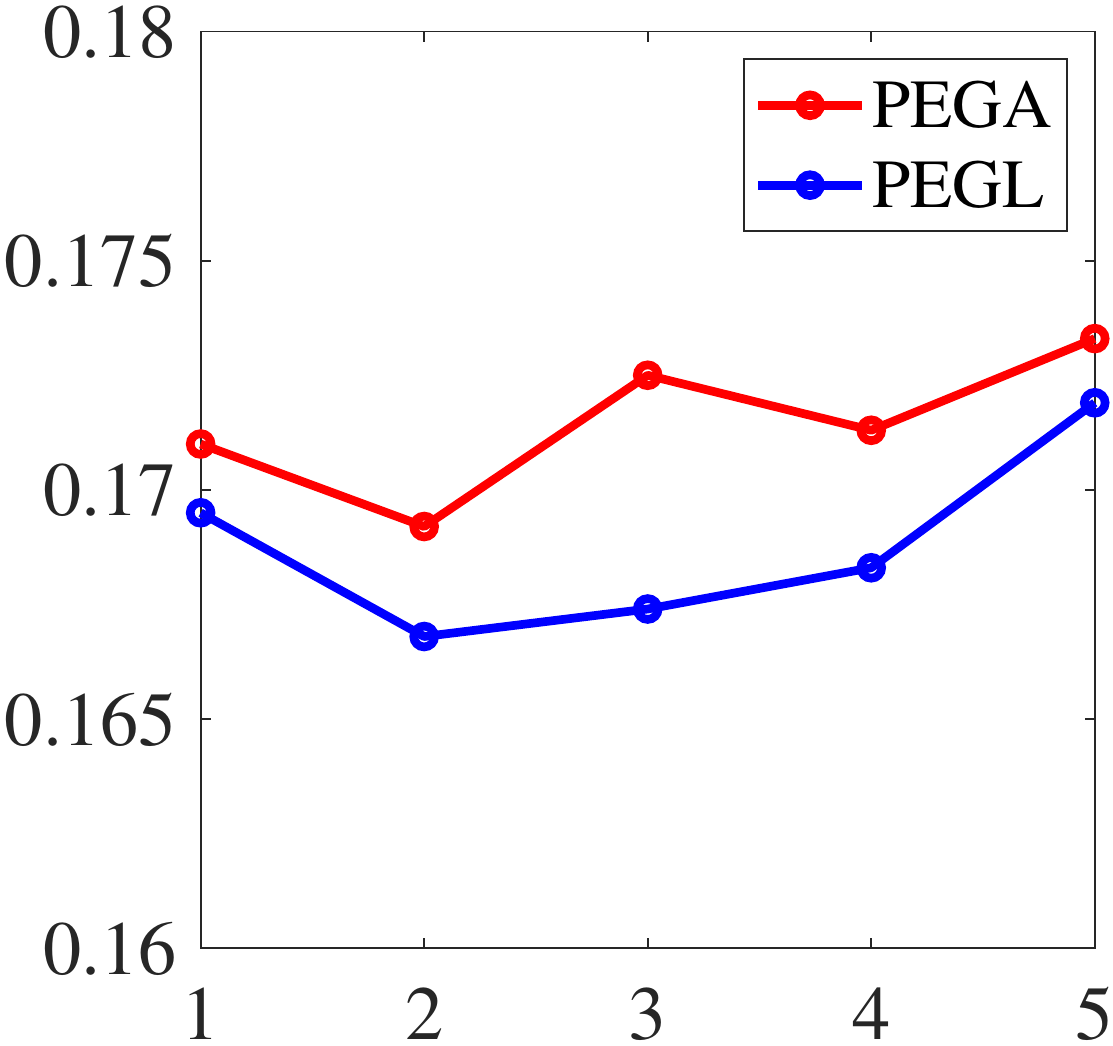}}
  \subfigure[$\tau$: length of short term]{
    \label{fig:t} 
    \includegraphics[width=0.31\columnwidth]{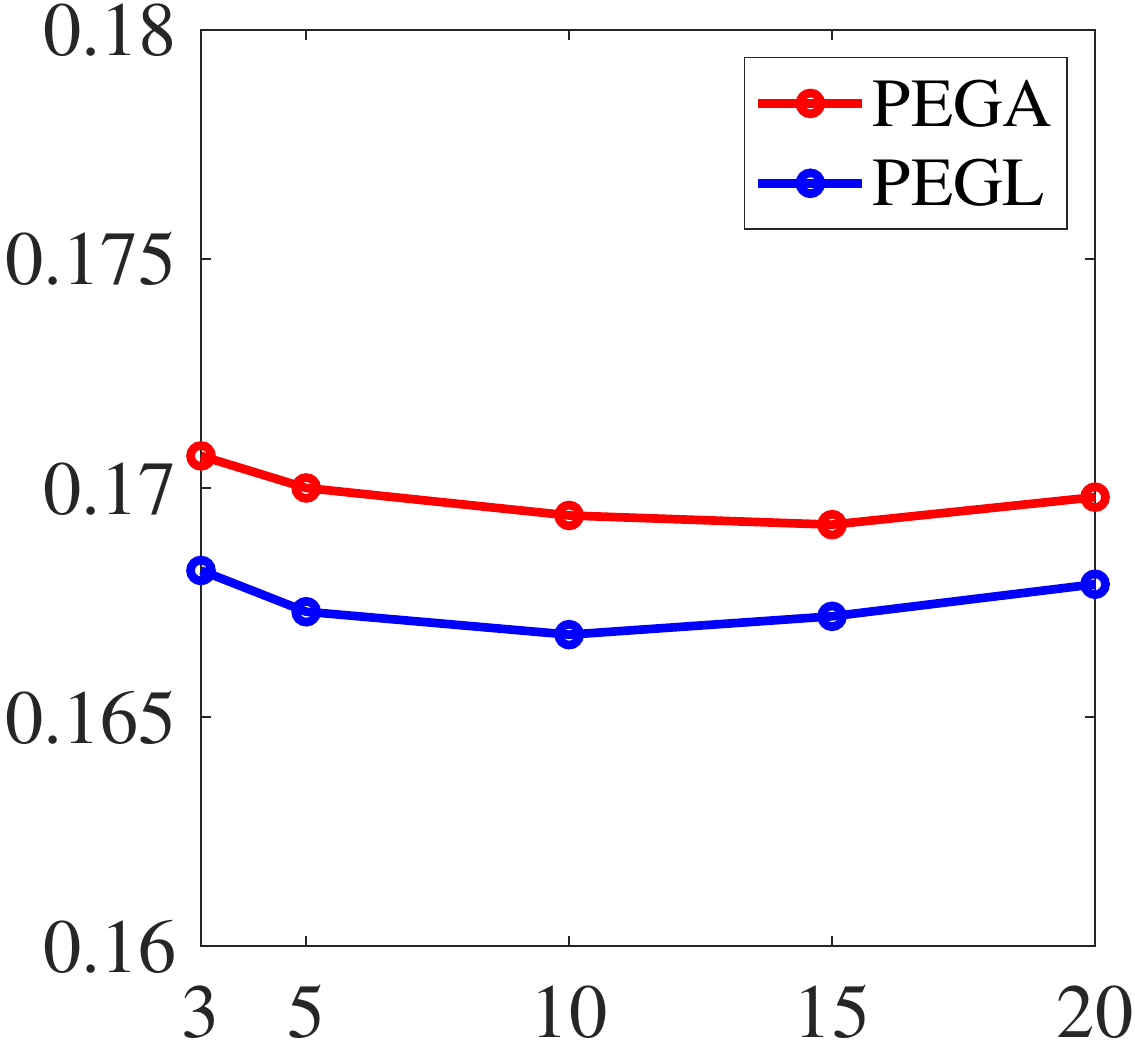}}
  \caption{The hyperparameter tune-up of MRSE performance on Sina Weibo, the observation time is 1.5h.}
  \label{fig:hyper} 
\end{figure}

Here we test the major hyperparameters $\{K, L, \tau\}$ of PEGA and PEGL and show the changes of their performance in Figure~\ref{fig:hyper}. Larger parameters do not necessarily lead to better results, and the most suitable hyperparameter of K, L, and $\tau$ is 4, 2, and 10 (or 15) on the Sina Weibo dataset. What's more, compared with K and L, the performance fluctuation caused by the change of $\tau$ is obviously small and stable. It may be the relatively slow evolution of preference that leads to this result.

\section{Conclusion}
In this paper, we contributed to the task of network-aware cascade size prediction and proposed a preference-enhanced graph model to increase the prediction accuracy. Our PEG model firstly developed a neural topic model to generate preference topics for the user sharing content. A preference shift modeling module was used for capturing the dynamic preferences of users to promote the social influence activating process. The experimental results demonstrated the superiority of PEG compared with state-of-the-art methods.

\section{Acknowledgments}
This research was partially supported by grants from National Natural Science Foundation of China (Grants No. U20A20229, 72101176). 

\bibliographystyle{ACM-Reference-Format}
\bibliography{FDGP}

\end{document}